# Hole spin dephasing time associated to hyperfine interaction in quantum dots

C. Testelin, F. Bernardot, B. Eble and M. Chamarro

*Institut des NanoSciences de Paris, Université P. et M. Curie, CNRS-UMR 7588,*

*140 rue de Lourmel, F-75015 Paris, France*

The spin interaction of a hole confined in a quantum dot with the surrounding nuclei is described in terms of an effective magnetic field. We show that, in contrast to the Fermi contact hyperfine interaction for conduction electrons, the dipole-dipole hyperfine interaction is anisotropic for a hole, for both pure or mixed hole states. We evaluate the coupling constants of the hole-nuclear interaction and demonstrate that they are only one order of magnitude smaller than the coupling constants of the electron-nuclear interaction. We also study, theoretically, the hole spin dephasing of an ensemble of quantum dots via the hyperfine interaction in the framework of frozen fluctuations of the nuclear field, in absence or in presence of an applied magnetic field. We also discuss experiments which could evidence the dipole-dipole hyperfine interaction and give information on hole mixing.

# I – INTRODUCTION

The spin of an individual electron, confined in a quantum dot (QD), is currently considered as a potential candidate for the realization of spintronic and quantum information processing in solid-state-based devices [1-3]. While in bulk or quantum wells, the electronic spin is efficiently relaxed by processes related to spin-orbit coupling, such as the D'Yakonov-Perel mechanism [4], the spatial confinement of carriers in semiconductor QDs significantly reduces the relaxation and decoherence processes. Recently, the hyperfine coupling with the spins of the lattice nuclei has been identified as the ultimate limit, at low temperature, to the electron spin relaxation or decoherence in QDs.

For conduction electrons, the hyperfine interaction has a Fermi contact character, and is at the origin of ensemble dephasing times of the order of one nanosecond in III-V QDs [5-9]. For holes, the Fermi contact coupling is massively suppressed because of the p-symmetry of the valence band states. The hyperfine interaction is then induced by the weaker long-range dipole-dipole coupling [10], so that much longer relaxation and decoherence times are expected [11].

Recent progresses in the preparation and reading of an ensemble of hole spins [12] or of a single hole spin, confined in QD, offer the opportunity to study their dynamics. By inserting single QDs in n-i-Schottky diode structures, D. Heiss *et al.*[13] and A.J. Ramsay *et al.*[14] have evidenced the possibility to initialize and store hole spins, as previously done with conduction electrons [15], while measuring the time dependence of their polarization.

In the present work, we show that, while being weaker than the electron Fermi contact interaction, the long-range dipole-dipole coupling between holes and nuclei can be an efficient decoherence mechanism, and leads to ensemble dephasing times of the order of ten nanoseconds in III-V QDs.

The paper is organized as follows: In section II, the hyperfine dipole-dipole coupling between nuclear spins and the spin of a hole is written in terms of an effective nuclear magnetic field acting on the hole spin. In this section, different hole states are considered: pure heavy-hole (hh), light-hole (lh) or mixed hole states. In section III, the hole spin dynamics in absence or in presence of an external magnetic field is calculated. In section IV, different experimental configurations are discussed.

## II – HYPERFINE INTERACTION OF A CONFINED VALENCE ELECTRON WITH NUCLEI

### A. Hyperfine coupling for a pure heavy or light hole

In contrast to electrons in the conduction band, the contact hyperfine interaction of a hole with nuclear spins is negligible, because holes in valence bands are described by p-symmetry Bloch functions. Then, the hyperfine interaction of nuclear spins with an electron in the valence band has a dipole-dipole nature. For a given nucleus, the Hamiltonian of this interaction writes [16]:

$$H_{dd}(\vec{I}) = 2\mu_B \frac{\mu_I}{I} \vec{I} \left[ \frac{\vec{\ell}}{\rho^3} - \frac{\vec{s}}{\rho^3} + 3\frac{\vec{\rho}(\vec{s}\cdot\vec{\rho})}{\rho^5} \right], \quad (1)$$

where $\mu_B$ is the Bohr magneton and $\mu_I$ is the nuclear magnetic moment; $\vec{I}$ is the nuclear spin operator; $\vec{\rho}$ is the electron position vector with origin at the nucleus position; $\vec{\ell} = \frac{\vec{\rho}\times\vec{p}}{\hbar}$ and $\vec{s}$ are the orbital momentum and spin operators, respectively.

A detailed calculation of the matrix elements of Hamiltonian $H_{dd}$, in the hh and lh bases, is presented in Appendix A. The hyperfine interaction being very small compared to the hh-lh splitting $\Delta_{lh}$, one can separate the 4 x 4 matrix $H_{dd}$ into two 2 x 2 matrices defined on the hh $\varphi_{\pm 3/2} = |J=3/2, J_z=\pm 3/2\rangle$ and lh $\varphi_{\pm 1/2} = |J=3/2, J_z=\pm 1/2\rangle$ bases (the z direction is aligned along the growth axis of the QD):

$$(H_{dd})_{hh} = \frac{8\mu_B\mu_I}{5I} |\Psi(\vec{R})|^2 \Omega \left\langle \frac{1}{\rho^3} \right\rangle \begin{pmatrix} I_z & 0 \\ 0 & -I_z \end{pmatrix}_{(\varphi_{+3/2},\varphi_{-3/2})}, \quad (2a)$$

$$(H_{dd})_{lh} = \frac{8\mu_B\mu_I}{15I} |\widehat{\Psi}(\vec{R})|^2 \Omega \left\langle \frac{1}{\rho^3} \right\rangle \begin{pmatrix} I_z & -2(I_x - iI_y) \\ -2(I_x + iI_y) & -I_z \end{pmatrix}_{(\varphi_{+1/2},\varphi_{-1/2})}, \quad (2b)$$

where $\Psi(\vec{R})$ and $\widehat{\Psi}(\vec{R})$ are the envelope functions of the $\varphi_{\pm 3/2}$ and $\varphi_{\pm 1/2}$ valence holes, respectively, taken at the nucleus position $\vec{R}$; $I_i$ ($i=x,y,z$) are the nuclear spin components; $\left\langle \frac{1}{\rho^3} \right\rangle$ is defined in Appendix A; $\Omega$ is the unit cell volume.

We underline here that the expressions (2 a) and (2 b) clearly show that the hole-nuclear hyperfine interaction is anisotropic in the nuclear spin components. For hh states, the hyperfine coupling is only induced by the nuclear spin components along the z-axis, while for lh states, this coupling arises from all the nuclear spin components, and mainly from the in-plane ones.

The absence of a valence electron, i.e. a hole, confined in a QD, interacts with a large number of nuclei. Then one has to consider the total hyperfine Hamiltonian:

$$H_{hf} = \sum_j H_{dd}(\vec{R}_j, \vec{I}^{\,j}), \quad (3)$$

where the summation runs over all the nuclei j, with position $\vec{R}_j$ and spin $\vec{I}^{\,j}$.

For hh states, this hyperfine interaction can be described in the hh basis by the effective Hamiltonian:

$$H_{hf}^h = \Omega \sum_j C_j |\Psi(\vec{R}_j)|^2 I_z^j S_z^h, \quad (4)$$

where $\vec{S}^h$ is a pseudo-spin with states $S_z^h = \pm 1/2$ associated to the hh $J_z = \pm 3/2$ states. The dipole-dipole hyperfine constants $C_j$ are defined as follows:

$$C_j = \frac{16}{5} \frac{\mu_B \mu_I^j}{I^j} \left\langle \frac{1}{\rho^3} \right\rangle_j. \quad (5)$$

It is then possible to define a nuclear field operator acting on the hh spin in a QD:

$$\vec{B}_N^h = \frac{\Omega}{g_h \mu_B} \sum_j C_j |\Psi(\vec{R}_j)|^2 I_z^j \vec{e}_z, \quad (6)$$

with $\vec{e}_z$ the unitary vector along the z direction, and $g_h$ the hh Landé factor; the right component of $g_h$ to be put in Eq. (6) is experimentally related to the direction of an external magnetic field (applied along a principal direction of the sample). The magnitude of this field, aligned along Oz, is randomly distributed from a QD to another QD, and the randomness is described by a 1D Gaussian probability density distribution:

$$P(B_{N_z}^h) = \frac{1}{\pi^{1/2} \Delta^h} \exp\left[-\frac{(B_{N_z}^h)^2}{(\Delta^h)^2}\right], \quad (7)$$

where $\Delta^h$ is the quadratic average of the nuclear field component, defined as:

$$(\Delta^h)^2 = 2\langle (B_{N_z}^h)^2 \rangle = \frac{2}{3}\left(\frac{\Omega}{g_h \mu_B}\right)^2 \sum_j I^j(I^j+1)(C_j)^2 |\Psi(\vec{R}_j)|^4. \quad (8)$$

As in ref. [5], this parameter can be related to $N_L$, the number of nuclei inside the QD:

$$\Delta^h = \frac{1}{g_h \mu_B} \sqrt{\frac{4 \sum_j I^j(I^j+1)(C_j)^2}{3 N_L}}, \quad (9)$$

where the summation is on all the nuclear species j.

For lh states, the hyperfine interaction is sensitive to all the nuclear components and can be written:

$$H_{hf}^l = \Omega \sum_j \frac{C_j}{3} |\hat{\Psi}(\vec{R}_j)|^2 \left[ -2I_x^j S_x^l - 2I_y^j S_y^l + I_z^j S_z^l \right]. \quad (10)$$

This leads to an effective nuclear field:

$$\vec{B}_N^l = \frac{\Omega}{g_l \mu_B} \sum_j \frac{C_j}{3} |\hat{\Psi}(\vec{R}_j)|^2 \left[ -2I_x^j \vec{e}_x - 2I_y^j \vec{e}_y + I_z^j \vec{e}_z \right], \quad (11)$$

where $\vec{e}_x$, $\vec{e}_y$ and $\vec{e}_z$ are the unitary vectors along the x, y and z directions, respectively; $g_l$ is the component of the lh Landé tensor corresponding to the direction of an applied magnetic field. Its magnitude and direction are described by the 3D Gaussian probability density distribution:

$$P(\vec{B}_N^l) = \frac{1}{\pi^{3/2} \Delta_{//}^2 \Delta_\perp} \exp\left[ -\frac{(B_{Nx}^l)^2}{\Delta_{//}^2} - \frac{(B_{Ny}^l)^2}{\Delta_{//}^2} - \frac{(B_{Nz}^l)^2}{\Delta_\perp^2} \right], \quad (12)$$

$$\Delta_{//}^2 = 4\Delta_\perp^2 = \frac{8}{3}\left(\frac{\Omega}{g_h \mu_B}\right)^2 \sum_j I^j(I^j+1)\left(\frac{C_j}{3}\right)^2 |\hat{\Psi}(\vec{R}_j)|^4 = \left(\frac{2}{3}\Delta^h\right)^2. \quad (13)$$

**B. Hyperfine coupling for mixed hole states**

**1 – Valence-band mixing**

Valence-band mixing arises from QD anisotropy, which can be induced by shape or strain. For instance, a symmetry reduction due to the confinement geometry of the dot, induces hole mixing through the off-diagonal terms of the Luttinger-Kohn Hamiltonian [17]. Another source of valence states mixing can be the absence of inversion symmetry in the growth direction, due to the dot shape or the intermixing chemical profile at the interfaces [18]. For flat and weakly elongated QDs, those contributions to valence mixing are expected small.

Several recent experiments have evidenced the mixed character of the hole states in self-assembled QDs. An efficient mixing between the hh and lh states can arise from the anisotropic relaxations of strain in the QDs [19-21]. While growing InAs on GaAs, InAs is compressed in the growth plane and distended in the growth direction; elastic and inelastic strain relaxation are involved in each QD. The Luttinger-Kohn Hamiltonian including strain effects [22] is given in Appendix B for a QD. Due to the spin-orbit interaction, it is adequate to limit the discussion to the states of angular momentum $J = 3/2$, defined in Appendix B. The confinement potential and the biaxial strain are responsible for a strong lift of degeneracy, noted $\Delta_{lh}$, between the valence band states $\varphi_{\pm 3/2}$ and $\varphi_{\pm 1/2}$. As in ref. [19], we will mainly consider the effects of strain anisotropy in the growth plane, and describe the strain on a QD by average values of $\varepsilon_{xy}$ and $\varepsilon_{xx} - \varepsilon_{yy}$ ($\varepsilon_{ij}$ denotes the $ij$ component of the strain tensor). In this approximation, the Hamiltonian can be written, in the $(\varphi_{+3/2}, \varphi_{-1/2}, \varphi_{-3/2}, \varphi_{+1/2})$ basis:

$$\begin{pmatrix} \Delta_{lh} & -R & 0 & 0 \\ -R^* & 0 & 0 & 0 \\ 0 & 0 & \Delta_{lh} & -R^* \\ 0 & 0 & -R & 0 \end{pmatrix} \qquad (14)$$

with $R = -\dfrac{\sqrt{3}}{2} b_v (\varepsilon_{xx} - \varepsilon_{yy}) + i d_v \varepsilon_{xy}$, where $b_v$ and $d_v$ are the deformation potentials for the valence band ($\varepsilon_{ij} = 0$ for $i$ or $j = z$).

The modified hh states can then be written:

$$\tilde{\varphi}_{+3/2} = \frac{1}{\sqrt{1+|\beta|^2}} (\varphi_{+3/2} + \beta\, \varphi_{-1/2}) \text{ and } \tilde{\varphi}_{-3/2} = \frac{1}{\sqrt{1+|\beta|^2}} (\varphi_{-3/2} + \beta^*\, \varphi_{+1/2}), \qquad (15)$$

with

$$\beta = |\beta| e^{i\phi} = i \frac{d_v \varepsilon_{xy}}{\Delta_{lh}} + \frac{\sqrt{3}}{2} \frac{b_v (\varepsilon_{xx} - \varepsilon_{yy})}{\Delta_{lh}}. \qquad (16)$$

In the following, we will limit our calculations to the first order in $\beta$, and will assume identical envelope functions for the hh and lh states ($\lambda(\vec{R}) = 1$ in Appendix A).

**2 – Hyperfine interaction for mixed hh-lh states**

From the results of the previous section, it is possible to show that the hyperfine interaction writes, in the basis of the mixed states $(\tilde{\varphi}_{+3/2}, \tilde{\varphi}_{-3/2})$:

$$H_{hf} = \Omega \sum_j C_j |\Psi(\vec{R}_j)|^2 \left[ \frac{2|\beta|}{\sqrt{3}} (\tilde{I}_x^j S_x + \tilde{I}_y^j S_y) + \tilde{I}_z^j S_z \right], \qquad (17)$$

with $\tilde{I}_x = \cos\phi\, I_x - \sin\phi\, I_y$, $\tilde{I}_y = \sin\phi\, I_x + \cos\phi\, I_y$ and $\tilde{I}_z = I_z$. $\vec{\tilde{I}}$ is thus obtained by a $-\phi$ rotation of $\vec{I}$ around z. Because of the z-rotation invariance of the nuclear field fluctuation distribution, changing $\vec{\tilde{I}}$ into $\vec{I}$ has no incidence on the dynamics of the QD ensemble.

The hyperfine interaction is anisotropic, either for pure or mixed hole states. We then propose to consider a general expression of the hyperfine Hamiltonian, which will be useful for pure or mixed states. Assuming an anisotropy factor $\alpha$, the hyperfine coupling can be written:

$$H_{hf} = \Omega \sum_j M_j |\Psi(\vec{R}_j)|^2 [\alpha(I_x^j S_x + I_y^j S_y) + I_z^j S_z] \qquad (18)$$

with $\alpha = 0$, $\alpha = 2$ (i.e. $\alpha = -2$, the ensemble dynamics is independent of the sign of $\alpha$) and $\alpha = 1$ for pure hh, lh and conduction electron, respectively. Small non-zero values of $\alpha = 2|\beta|/\sqrt{3}$ will be associated to mixed hh-lh states. The hyperfine constants are $M_j = A_j$ (ref. [5]), $M_j = C_j$ or $M_j = C_j/3$ for the conduction electron, the pure hh and lh states, respectively; $M_j = C_j$ for the mixed hh-lh states of Eqs. (15).

The considered spin is then submitted to an effective nuclear field:

$$\vec{B}_N = \frac{\Omega}{g\mu_B} \sum_j M_j |\Psi(\vec{R}_j)|^2 [\alpha(I_x^j \vec{e}_x + I_y^j \vec{e}_y) + I_z^j \vec{e}_z] \qquad (19)$$

with $g$ the Landé factor of the considered state. Once again, the nuclear field components are assumed to follow a Gaussian distribution:

$$P(\vec{B}_N) = \frac{1}{\pi^{3/2} \alpha^2 \Delta_0^3} \exp\left[-\frac{(B_{Nx})^2}{\alpha^2 \Delta_0^2} - \frac{(B_{N_y})^2}{\alpha^2 \Delta_0^2} - \frac{(B_{Nz})^2}{\Delta_0^2}\right], \qquad (20)$$

with

$$\Delta_0^2 = \frac{2}{3}\left(\frac{\Omega}{g\mu_B}\right)^2 \sum_j I^j(I^j+1)(M_j)^2 |\Psi(\vec{R}_j)|^4. \qquad (21)$$

In Table I [23-26], we have reported, for different atomic species, values of the hyperfine constants: $A_j$ (as defined in ref. [5]) for a conduction electron and $C_j$ for a hh. To estimate them, we have used calculated and measured values of the parameters $|u_c(\vec{0})|$ (value of the conduction Bloch function at the nucleus position) and $\left\langle \frac{1}{\rho^3} \right\rangle$. The last column of Table I gives the natural concentration of isotopes carrying a non-zero nuclear spin. The main information is that the hyperfine constant for hh is typically one order of magnitude smaller than for electrons; this differs from the common hypothesis that the hole-nuclear interaction is fully negligible, and is in agreement with recent calculations of Fisher *et al* [11]. One can also observe that, for a given carrier, the coupling constants are comparable for III-V and II-VI compounds, so that the amplitude of the hyperfine coupling with all the QD nuclei mainly depends on the isotope distribution and the QD size.

In the next section, we discuss the hole spin dynamics in absence or in presence of an applied magnetic field. In the first case, we will center our discussion on the influence of the anisotropy of the interaction, and in the second case, different configurations of the applied magnetic field will be considered.

## III – HOLE SPIN DYNAMICS AND DEPHASING FOR ENSEMBLES OF QDs

### A. Hole spin dephasing in a fluctuating nuclear field

To study the time dependence of an ensemble of hole spins, one can follow the approach developed by Merkulov *et al.* [5] for an isotropic hyperfine interaction between nuclei and conduction electrons. We neglect the nuclear dipole-dipole interactions, which do not conserve the total spin of the hole-nuclear system. These interactions become important only at times longer than $10^{-4}$ s.

Let us consider an ensemble of identical QDs containing a single hole, all prepared with the same initial spin $\vec{S}_0$. Due to the randomly oriented nuclear spins, the nuclear hyperfine fields inside the dots differ from QD to QD, and have a different effect on the initial hole spin.

As in ref. [5], we consider the time dependence of the ensemble average hole spin relaxation for times small compared to the period of the nuclear precession in the hyperfine field of a hole (approximation of the frozen nuclear field fluctuations). In each QD, the hole spin precesses in a total magnetic field $\vec{B}+\vec{B}_N$, with $\vec{B}$ an applied magnetic field. Because of the anisotropy of the nuclear field distribution, two cases are of particular interest: $\vec{B}$ along the z-axis (the system stays invariant by rotation around z-axis), and $\vec{B}$ in the xy plane (so that any rotation invariance disappears). For an ensemble of spins in the initial state $\vec{S}_0 = S_{0x}\vec{e}_x + S_{0y}\vec{e}_y + S_{0z}\vec{e}_z$, the time-dependent average spin $\langle \vec{S}(t) \rangle$ can be deduced from the precession of $\vec{S}_0$ in the random field $\vec{B}+\vec{B}_N$ within each QD, and is written:

$$\langle \vec{S}(t) \rangle = [S_{0x}R_x(t) - S_{0y}R_y(t)]\vec{e}_x + [S_{0x}R_y(t) + S_{0y}R_x(t)]\vec{e}_y + S_{0z}R_z(t)\vec{e}_z \quad \text{for } \vec{B}//\vec{z} \quad (22a)$$

$$\langle \vec{S}(t) \rangle = S_{0x}R_x^1(t)\vec{e}_x + [S_{0y}R_y^2(t) + S_{0z}R_y^1(t)]\vec{e}_y + [-S_{0y}R_z^2(t) + S_{0z}R_z^1(t)]\vec{e}_z \quad \text{for } \vec{B}//\vec{x} \quad (22b)$$

The expressions of $R_i$ and $R_i^j$ ($i = x, y, z$ ; $j = 1, 2$) are given in Appendix C.

**B - Hole spin dynamics in zero magnetic field – influence of the anisotropy**

In absence of applied magnetic field, the average hole spin is reduced to:

$$\left\langle \vec{S}(t) \right\rangle = S_{0x} R_x(t) \vec{e}_x + S_{0y} R_x(t) \vec{e}_y + S_{0z} R_z(t) \vec{e}_z. \quad (23)$$

As for the case of an isotropic hyperfine interaction (case of conduction electrons), we can define an ensemble dephasing time $T_{\Delta_0}$ from the coupling constants $M_j$:

$$T_{\Delta_0} = \frac{\hbar}{g\mu_B \Delta_0} = \hbar \sqrt{\frac{3N_L}{4\sum_j I^j(I^j+1)M_j^2}} \quad (24)$$

Table II [27-29] gives an overview of the estimated dephasing times for electrons and hh, for the most usually studied III-V and II-VI QDs. We underline that dephasing times of the II-VI compounds are 3-10 times larger than dephasing times for III-V compounds, due to the very low natural abundance of isotopes with non-zero nuclear magnetic moment (see Table I).

In the following, we will study the average hole spin dynamics and will use scaling laws by taking the normalized magnetic field $\delta = \frac{B}{\Delta_0}$ and the normalized time $\tau = \frac{t}{T_{\Delta_0}}$. Figure 1 shows the time dependence of the $R_z(\tau)$ and $R_x(\tau)$ components of $\left\langle \vec{S}(t) \right\rangle$ (see Eq. (23)), for anisotropy factors $\alpha$ varying from $\alpha = 0$ (pure hh state) to $\alpha = 2$ (pure lh state).

For conduction electrons ($\alpha=1$), we retrieve the result of ref. [5], the ensemble average spin polarization decreases and shows two regimes: the first regime consists of an initial fall of the spin polarization, which makes it reach 4% of its initial value within a characteristic time $2T_{\Delta_0}$; the second regime is a plateau of the spin polarization, at 1/3 of its initial value, reached from a typical time of $4T_{\Delta_0}$.

Figure 1 also shows that $R_z(\tau)$ and $R_x(\tau)$ present the same general behaviour described for an isotropic interaction, for an anisotropy factor $\alpha \neq 0$. The minimum is close to $t = \frac{2T_{\Delta_0}}{\alpha}$, and the value of this minimum of polarization depends on the observed component and on the value of α. After a fast decrease of the spin components, one reaches a steady-state value for the x and z components, with $R_x(\infty) > R_z(\infty)$ for 1<α, and $R_x(\infty) < R_z(\infty)$ for α<1. Figure 2 gives the steady-state values for these two components, as a function of the anisotropy factor.

$R_z(\infty)$ decreases from 1 to zero when α increases from zero to α>>1, while $R_x(\infty)$ increases from zero to one half.

The dynamics of the average spin polarization is very different for pure hh states ($\alpha = 0$) prepared in the eigenstate $S_z^h = \pm 1/2$ ($J_z^h = \pm 3/2$). The randomly fluctuating nuclear hyperfine field, aligned with z, has then no influence on the average hh spin and no dephasing occurs: $R_z(\tau)$ keeps constant and equals to one, see black lines in Figure 1. For an in-plane spin component, one observes a Gaussian time dependence of the form $R_x(t) = e^{-\left(\frac{t}{2T_{\Delta_0}}\right)^2}$, reaching zero for $t >> T_{\Delta_0^h}$, as already discussed in Eq. (11) of ref [11].

## C - Hole spin dynamics in presence of an applied magnetic field

In the following, for clarity, we limit ourselves to the cases where $\vec{B}$ (applied field) and $\vec{S}_0$ are in the zx plane ($B_y = 0$ and $S_{0y} = 0$).

## C1 Case $\vec{B} // z$

Figure 3 shows the time dependence of the different components of $\langle \vec{S}(t) \rangle$ which appear in Eq.(22a), for pure hh (α=0) and lh (α=2) spins. The curves have been plotted versus the reduced time $\tau = t/T_{\Delta_0}$, and have been calculated for different values of the reduced magnetic field $\delta = B/\Delta_0$. The upper curves of Figure 4 show also the time dependence of the $R_x(\tau)$, $R_y(\tau)$ and $R_z(\tau)$ components for a mixed heavy-light hole (α=0.5). In presence of an applied magnetic field, along z, all the spin components tend to a steady-state value after several oscillations. With increasing magnetic field, the frequency of oscillation increases, and the steady state value tends to 1 for the longitudinal spin component $R_z(\infty)$ and reaches zero for the transverse components $R_x(\infty)$ and $R_y(\infty)$. In a general way, for an $\alpha \neq 0$, the behaviour of $R_x(\tau)$, $R_y(\tau)$ and $R_z(\tau)$ follows the general trends obtained by Merkulov *et al* [5] for conduction electrons. For a pure hh (α=0), however, a very different behaviour is observed; notably, for any field $R_z(\tau)=1$ and $R_x(\infty)=0$.

In the high-field limit ($\delta >> 1$), the spin components can be written, in the second order in $\delta^{-1}$:

$$R_x(t) = \frac{\alpha^2}{2\delta^2} + \left\{\left[1 - \frac{\alpha^2}{2\delta^2} - \frac{\alpha^2(\alpha^2-1)}{\delta^2}\left(\frac{t}{2T_{\Delta_0}}\right)^2\right]\cos\omega_B t - \frac{\alpha^2}{\delta}\frac{t}{2T_{\Delta_0}}\sin\omega_B t\right\}e^{-\left(\frac{t}{2T_{\Delta_0}}\right)^2}, \quad (25a)$$

$$R_y(t) = \left\{\left[1 - \frac{\alpha^2}{2\delta^2} - \frac{\alpha^2(\alpha^2-1)}{\delta^2}\left(\frac{t}{2T_{\Delta_0}}\right)^2\right]\sin\omega_B t + \frac{\alpha^2}{\delta}\frac{t}{2T_{\Delta_0}}\cos\omega_B t\right\}e^{-\left(\frac{t}{2T_{\Delta_0}}\right)^2}, \quad (25b)$$

$$R_z(t) = 1 - \alpha^2 \frac{1 - e^{-\left(\frac{t}{2T_{\Delta_0}}\right)^2}\cos\omega_B t}{\delta^2}, \quad (25c)$$

where $\omega_B = \frac{g\mu_B B}{\hbar}$ is the Larmor precession frequency induced by the applied magnetic field. These expressions clearly show the Gaussian time dependence of the spin components, with a dephasing time $T_{\Delta_0}$, and the field dependence of the steady-state values.

**C2. Case $\vec{B}//x$**

For an in-plane magnetic field, in the high-field limit ($\delta \gg 1$), the spin components can be written, in the second order in $\delta^{-1}$ ($\alpha \neq 0$):

$$R_x^1(t) = 1 - (\alpha^2 + 1)\frac{1 - e^{-\left(\frac{\alpha t}{2T_{\Delta_0}}\right)^2}\cos\omega_B t}{2\delta^2}, \quad (26a)$$

$$R_y^1(t) = \left\{\left[1 - \frac{\alpha^2+1}{4\delta^2} + \frac{\alpha^4+2\alpha^2-3}{8\delta^2}\left(\frac{t}{2T_{\Delta_0}}\right)^2\right]\sin\omega_B t + \frac{\alpha^2+1}{2\delta}\frac{t}{2T_{\Delta_0}}\cos\omega_B t\right\}e^{-\left(\frac{\alpha t}{2T_{\Delta_0}}\right)^2}, \quad (26b)$$

$$R_z^1(t) = \frac{1}{2\delta^2} + \left\{\left[1 - \frac{1}{2\delta^2} + \frac{\alpha^4+2\alpha^2-3}{\delta^2}\left(\frac{t}{2T_{\Delta_0}}\right)^2\right]\cos\omega_B t - \frac{\alpha^2+1}{2\delta}\frac{t}{2T_{\Delta_0}}\sin\omega_B t\right\}e^{-\left(\frac{\alpha t}{2T_{\Delta_0}}\right)^2}. \quad (26c)$$

One can then clearly evidence a dephasing time $T_{//} = \frac{T_{\Delta_0}}{\alpha}$, with a minimum value for pure lh.

For pure hh states ($\alpha = 0$), the previous high-field expressions are not valid. For $\alpha = 0$, the time-dependence of the spin components is totally different, as already mentioned in ref. [11] for $R_z^1(t)$. The spin components are then given, in the strong-field regime, by the following expressions:

$$R_x^1(t) = 1 - \frac{1}{2\delta^2} + \frac{1}{2\delta^2} \frac{\cos\left(\omega_B t + \frac{3}{2}\arctan\frac{t}{\tau_{//}}\right)}{\left(1+\left(\frac{t}{\tau_{//}}\right)^2\right)^{3/4}} \quad , \quad (27a)$$

$$R_y^1(t) = \frac{\sin\left(\omega_B t + \frac{1}{2}\arctan\frac{t}{\tau_{//}}\right)}{\left(1+\left(\frac{t}{\tau_{//}}\right)^2\right)^{1/4}} - \frac{1}{4\delta^2} \frac{\sin\left(\omega_B t + \frac{3}{2}\arctan\frac{t}{\tau_{//}}\right)}{\left(1+\left(\frac{t}{\tau_{//}}\right)^2\right)^{3/4}} \quad , \quad (27b),$$

$$R_z^1(t) = \frac{\cos\left(\omega_B t + \frac{1}{2}\arctan\frac{t}{\tau_{//}}\right)}{\left(1+\left(\frac{t}{\tau_{//}}\right)^2\right)^{1/4}} + \frac{1}{2\delta^2} - \frac{1}{2\delta^2} \frac{\cos\left(\omega_B t + \frac{3}{2}\arctan\frac{t}{\tau_{//}}\right)}{\left(1+\left(\frac{t}{\tau_{//}}\right)^2\right)^{3/4}} \quad . \quad (27c)$$

A new dephasing time which depends on the value of the applied magnetic field is defined, $\tau_{//} = 2\delta T_{\Delta_0} = 2\frac{B}{\Delta_0}T_{\Delta_0}$, while the longitudinal and transverse components decrease as $t^{-3/2}$ and $t^{-1/2}$, respectively. As observed in Figure 5 for pure hh in zero magnetic field, the spin z-component is constant, $R_z^1(t) = 1$, while an in-plane magnetic field induces dephasing and reduces the mean value of $R_z^1(t)$ which shows an oscillatory pattern. For $R_x^1(t)$, a high field is then necessary to reach a steady-state regime where $R_x^1(t)$ becomes close to 1. This behavior can be easily understood: (i) if $S_{ox} = S_{oy} = 0$ and $S_{oz} \neq 0$, in zero magnetic field, the hh spins and nuclear fields are aligned, so that no dephasing can occur, whatever the magnitude of the nuclear field fluctuations; (ii) in presence of a small magnetic field, of the order of the typical nuclear field fluctuation $\Delta_0$, the hh spins precess around total magnetic fields out of the z-axis, so that the spin components are sensitive to the nuclear field fluctuations, and a decrease of the average spin amplitude then occurs; (iii) in a strong magnetic field, the hyperfine nuclear field is screened, so that the dephasing time $\tau_{//}$ increases and for finite values of $\tau$, the $R_x^1(t)$ spin amplitude tends to 1: in this regime, an initial ($S_{ox} \neq 0, S_{oy} = S_{oz} = 0$) spin essentially remains constant in time.

Finally, let us compare the time dependence of the transverse components ($R_x^1$ and $R_y^1$) for pure hh or lh spins, at high field, shown in Figure 5 and 6. One clearly observes a Gaussian decay of the oscillations for lh ($\alpha = 2$) spins, and a power-law decay for hh ($\alpha = 0$) spins. The lower part of Figure 4 shows the behaviour of $R_x^1(\tau)$, $R_y^1(\tau)$ and $R_z^1(\tau)$ for a mixed heavy-light hole spin. These components follow the general trends already given for the case of a lh spin ($\alpha = 2$).

In the last section we will connect the above commented theoretical results with expected experimental observations.

## IV DISCUSSION

A very useful tool to experimentally study the electron spin polarization is the analysis and the measurement of the degree of circular polarization of the photoluminescence of samples containing p-doped QDs. In this case, after a non-resonant optical excitation and the subsequent relaxation of the photo-created electron-hole pairs, a positively-charged trion is created in some QDs. This photo-created species contains three particles: two antiparallel holes and one electron with its spin pointing up or down depending on the helicity of the circular polarization of the exciting light. Braun *et al.* [8] have evidenced that the hyperfine interaction is at the origin of the electron spin dephasing in self-assembled QDs by analysing of the temporal behaviour of the degree of polarization of positively charged trions in PL experiments.

By analogy, one could think that the study of the decay of the PL degree of polarization of a n-doped sample containing QDs would give information on the hole spin dynamics and dephasing. However, experimental constraints for n-doped samples are slightly different and make the final task much more difficult. The main experimental difference is given by the fact that the lifetime of photo-created trions is in the order of 1 ns, and during this time the electron spin evolves with a dephasing time in the order of 500ps; for holes, a much longer dephasing time is expected, and then no significant evolution during lifetime should be observed, as confirmed by several experimental studies[30,31].

To get experimental information on the hole spin dynamics, pump-probe experiments on samples containing p-doped QDs are more appropriate, such as the measurement of the photo-induced Faraday or Kerr rotation [32,33], or of the photo-induced circular dichroism [27, 34]. In these experiments an initial hole spin polarization is created by a resonant excitation of charged trions and subsequent transfer of their spin polarization to the hole spin. The observed

Faraday or Kerr rotation is related to the component of the spin polarization along the light propagation direction (z direction).

Another recent possibility consists to initialize a single hole spin in a QD immerged in a diode structure, using a resonant optical excitation of an electron-hole pair in the QD followed by a fast electron tunneling controlled by a applied voltage. The readout of the hole spin state is then obtained by measuring the photocurrent through the diode under spin –selective optical excitation of trions [14].

Usually two main configurations are considered to study the action of a magnetic field upon a phenomenon, as for exemple here the dynamics of the average spin: the Faraday configuration and the Voigt configuration. In the first one the magnetic field is applied parallel to the direction of the previously optically created spin, meanwhile in the second one, the magnetic field is applied perpendicular to the photo-created spin.

Figure 7 (a) shows the behaviour of the steady-state amplitude of $R_z(t >> T_{\Delta_0})$ as a function of the normalized magnitude of a Faraday magnetic field, $\delta = B/\Delta_0$. We observe that the effect of an external field, B, applied along z direction is very important for the conservation of the initial hole spin, i.e. for the quenching of the hyperfine effect on the $R_z$ component. A very small field, of the order of several $\Delta_0$ (i.e. a few mT), suppresses the relaxation of the longitudinal hole spin component, $R_z$, when $\alpha \neq 0$. A quasi-Lorenzt curve is obtained, its amplitude is mainly fixed by the $\alpha$ factor and its line-width is given by the hyperfine interaction coupling strength $T_\Delta$ for a known longitudinal Landé factor, $g^z$. This behaviour has been recently observed in p-doped InAs/GaAs QDs [12].

Figure 7 (b) shows the magnetic field dependence of the steady-state amplitude of $R_z^1(t >> T_{\Delta_0})$ for a Voigt configuration. The reduced value of the magnetic field, $\delta$, is used. Once again a very small magnetic field has an important effect on the value of the z-component of the average spin. The main difference, here, with respect to the Faraday configuration is that $\alpha = 0$ does not present a singular behaviour. The amplitude of all curves decreases to zero (whatever the value of $\alpha$), following a curve for which the amplitude is mainly fixed by the anisotropy factor $\alpha$ and the half-width which is only function of the hyperfine interaction coupling $T_\Delta$ when the transverse Lande factor, $g^x$, is known.

Then, from the experimental study of the magnetic field dependence of the steady-state value of the average hole spin polarization in Faraday and/or Voigt configuration it is possible to obtain information on the hole spin dephasing time and on the degree of purity or mixing of hole states as shown in Figure 7 a) and b).

**CONCLUSION**.

We have calculated the hole-nuclear hyperfine interaction in QDs for pure hh or lh and for mixed heavy-light holes, and its consequences in the hole spin dynamics of an ensemble of QDs. In contrast to the electron-nuclear hyperfine interaction in QDs, the hole-nuclear hyperfine interaction is highly anisotropic for a pure hh, this anisotropy being reduced by hh-lh mixing. We have shown also that contrary to the common idea the hole hyperfine interaction is far from negligible because the long range dipole-dipole term induces a coupling which is only one order of magnitude smaller than the electron-nuclear interaction. This result has an important effect on the hole spin dynamics and its potential use as a quantum bit, since the hyperfine interaction is the main source of decoherence, at low temperature, for holes confined in a QD.

Finally, a first criterium to reduce decoherence in III-V compounds is to obtain isotropic and strainless QDs. In this case, the hole spin is in a pure hh state and, for a hole spin polarized along z the growth axis, all decoherence phenomena induced by hyperfine coupling are minimized or suppressed. Nonetheless, decoherence processes are still possible for transverse spin components. Due to the small value of $\Delta_0$, the dispersion of the nuclear field distribution, a very small magnetic field can be used to screen the hole hyperfine coupling. Another possibility to reduce decoherence is to consider II-VI QDs, with a majority of non-magnetic nuclei, the hyperfine coupling being totally canceled in a QD made exclusively with isotopes without nuclear spin.

ACKNOWLEDGEMENTS

We acknowledge O. Krebs, T. Amand and X. Marie for fruitful discussions. One of us (B.E.) thanks the C'Nano-IdF for its financial support.

**APPENDIX A**

The hyperfine interaction of a nuclear spin with an electron in the valence band has a dipole-dipole nature and can be written [16]:

$$H_{dd}(\vec{I}) = 2\mu_B \frac{\mu_I}{I} \vec{I} \left[ \frac{\vec{\ell}}{\rho^3} - \frac{\vec{s}}{\rho^3} + 3\frac{\vec{\rho}(\vec{s}.\vec{\rho})}{\rho^5} \right] \quad \text{(A-1)}$$

where $\mu_B$ is the Bohr magneton and $\mu_I$ is the nuclear magnetic moment; $\vec{I}$ is the nuclear spin operator; $\vec{\ell} = \frac{\vec{\rho} \times \vec{p}}{\hbar}$ and $\vec{s}$ are the orbital momentum and spin operators, respectively, and $\vec{\rho}$ is the electron position vector with origin at the nucleus position.

Using notations similar to those of the ref. [10], the dipole-dipole Hamiltonian can be written:

$$H_{dd}(\vec{I}) = 2\mu_B \frac{\mu_I}{I} (V_1 - V_2) \quad \text{(A-2)}$$

where $V_1 = \frac{\ell_m}{\rho^3} I_m$, $V_2 = P_{mn}(\vec{\rho}) s_m I_n$ and $P_{mn}(\vec{\rho}) = \frac{\rho^2 \delta_{mn} - 3\rho_m \rho_n}{\rho^5}$, with $m, n = x, y, z$. In the expression of $H_{dd}$, the summation is done over all possible values of $m$ and $n$.

The electron wave function is defined as $\varphi(\vec{r}) = \sqrt{\Omega} \, \Psi(\vec{r}) \, u(\vec{r})$, where $u(\vec{r})$ is the Bloch function, normalized on a unit cell of volume $\Omega$, and $\Psi(\vec{r})$ is the quantum dot envelope wave function, normalized on the sample volume ($\vec{r}$ is the space position vector). $\Psi(\vec{r})$ is related to its Fourier transform $\phi(\vec{k})$ by the relation:

$$\Psi(\vec{r}) = \frac{1}{(2\pi)^{3/2}} \int \Phi(\vec{k}) \, e^{i\vec{k}\vec{r}} d\vec{k} \quad \text{(A-3)}$$

We consider the calculation of the matrix elements of $H_{dd}$, in the basis formed by the valence band states $\varphi_{\pm 3/2} = |J = 3/2, J_z = \pm 3/2\rangle$ and $\varphi_{\pm 1/2} = |J = 3/2, J_z = \pm 1/2\rangle$. First, this leads to the calculation of integrals of the form:

$$\tilde{Q}_{ijmn} = \Omega \int d\vec{\rho} \, F_i(\vec{\rho}) \, F_j(\vec{\rho}) \, P_{mn}(\vec{\rho}) \, \Psi^*(\vec{R}+\vec{\rho}) \, \Psi(\vec{R}+\vec{\rho}) \quad \text{(A-4a)}$$

$$\tilde{T}_{mij} = \frac{\Omega}{\hbar} \int d\vec{\rho} \, F_i(\vec{\rho}) \, \Psi^*(\vec{R}+\vec{\rho}) \, \frac{|\vec{\rho} \wedge \vec{p}|_m}{\rho^3} F_j(\vec{\rho}) \, \Psi(\vec{R}+\vec{\rho}) \quad \text{(A-4b)}$$

$\vec{\rho} = \vec{0}$ corresponds to the position of the nucleus under study, located at $\vec{r} = \vec{R}$. $F_i(\vec{\rho})$ are the orbital functions of p-symmetry, $|X\rangle$, $|Y\rangle$ and $|Z\rangle$ for $i = x$, $y$ and $z$ respectively. These integrals can be rewritten:

$$\tilde{Q}_{ijmn} = \frac{\Omega}{(2\pi)^3} \int d\vec{K}.d\vec{q} \, \Phi^*(\vec{K} - \frac{\vec{q}}{2}) \, \Phi(\vec{K} + \frac{\vec{q}}{2}) \, e^{i\vec{q}.\vec{R}} \, Q_{ijmn} \quad \text{(A-5a)}$$

$$\tilde{T}_{ijmn} = \frac{\Omega}{(2\pi)^3} \int d\vec{K}.d\vec{q}\, \Phi^*(\vec{K}-\frac{\vec{q}}{2})\, \Phi(\vec{K}+\frac{\vec{q}}{2})\, e^{i\vec{q}.\vec{R}}\, T_{ijmn} \qquad (A\text{-}5b)$$

To calculate those quantities, it is convenient to first calculate the integrals:

$$Q_{ijmn} = \int d\vec{\rho}\, F_i(\vec{\rho})\, F_j(\vec{\rho})\, P_{mn}(\vec{\rho})\, e^{i\vec{q}.\vec{\rho}} \qquad (A\text{-}6a)$$

$$T_{mij} = \frac{1}{\hbar} \int d\vec{r}\, F_i(\vec{\rho})\, e^{i\vec{q}.\vec{\rho}}\, \frac{|\vec{\rho} \wedge \vec{p}|_m}{\rho^3}\, F_j(\vec{\rho}) \qquad (A\text{-}6b)$$

As described in ref. [10], one can enclose the nucleus in a sphere of radius $R_0$ so that the inequalities $R_0 \gg a_0$ and $qR_0 \ll 1$ are simultaneously satisfied. This leads to the relations:

$$Q_{ijmn} = A_{ijmn} + B_{ijmn} \quad \text{and} \quad T_{mij} = iC_{mij} + iD_{mij} \qquad (A\text{-}7)$$

The tensors $A$ and $D$ are related to integrals over the cells corresponding to $r < R_0$, and $B$ and $C$ to integrals over the cells corresponding to $r > R_0$ (similar expression can be defined for $\tilde{Q}_{ijmn} = \tilde{A}_{ijmn} + \tilde{B}_{ijmn}$ and $\tilde{T}_{mij} = i\tilde{C}_{mij} + i\tilde{D}_{mij}$). These tensors are given by E.I. Gryncharova *et al.* [10]:

$$A_{ijmn} = \frac{2}{5}\left\langle \frac{1}{\rho^3} \right\rangle \left[ \delta_{ij} - \frac{3}{2}(\delta_{im}\delta_{jn} + \delta_{in}\delta_{jm}) \right] \qquad (A\text{-}8a)$$

$$B_{ijmn} = -\frac{\delta_{ij}}{\Omega}\frac{16\pi}{3\sqrt{2}}\left( \delta_{mn} - 3\frac{q_m q_n}{q^2} \right) \qquad (A\text{-}8b)$$

$$C_{mij} = \frac{1}{\Omega}\frac{16\pi}{3\sqrt{2}}\left[ \varepsilon_{mij} + \frac{3}{2}\frac{(\varepsilon_{mni}q_n q_j - \varepsilon_{mnj}q_n q_i)}{q^2} \right] \qquad (A\text{-}8c)$$

$$D_{mij} = -\varepsilon_{mij}\left\langle \frac{1}{\rho^3} \right\rangle \qquad (A\text{-}8d)$$

with $\left\langle \frac{1}{\rho^3} \right\rangle = \int_\Omega d\vec{\rho}\, X^2(\vec{\rho})\, \frac{1}{\rho^3}$. $\varepsilon_{mij}$ is the unit antisymmetric tensor of rank three.

The short-range contributions $\tilde{A}_{ijmn}$ and $\tilde{D}_{mij}$ to integrals $\tilde{Q}_{ijmn}$ and $\tilde{T}_{mij}$, respectively, can be deduced easily:

$$\tilde{A}_{ijmn} = \Omega\, A_{ijmn}\, |\Psi(\vec{R})|^2 \quad \text{and} \quad \tilde{D}_{mij} = \Omega\, D_{mij}\, |\Psi(\vec{R})|^2 \qquad (A\text{-}9)$$

The long range contributions $\tilde{B}_{ijmn}$ and $\tilde{C}_{mij}$ are more complicated integrals. Nonetheless, one can estimate their order of magnitude, by assuming a Gaussian envelope function $\Psi(\vec{r}) = \frac{1}{\pi^{3/4}a^{3/2}}e^{-r^2/2a^2}$ for simplicity. One can then calculate :

$$\tilde{B}_{ijmn} = -\delta_{ij}\frac{16\pi}{3\sqrt{2}}\left( \delta_{mn} - 3\frac{R_m R_n}{R_p^2} \right)|\Psi(\vec{R})|^2 \qquad (A\text{-}10a)$$

$$\tilde{C}_{mij} = \frac{16\pi}{3\sqrt{2}}\left( \varepsilon_{mij} - \frac{3}{2}\frac{\varepsilon_{mni}R_n R_j - \varepsilon_{mnj}R_n R_i}{R_p^2} \right)|\Psi(\vec{R})|^2 \qquad (A\text{-}10b)$$

the $R_i$ ($i = x, y, z$) being the components of vector $\vec{R}$. The ratio between the short- and long-range contributions is of the order of $\sigma = \frac{\sqrt{2}}{16\pi}\Omega\left\langle\frac{1}{\rho^3}\right\rangle$. For InAs and GaAs compounds, $\sigma \approx 30-60$, so that the long-range contributions $\tilde{B}_{ijmn}$ and $\tilde{C}_{mij}$ can be neglected.

Finally, the dipole-dipole Hamiltonian, between a valence electron and a nucleus at position $\vec{R}$ and with nuclear spin $\vec{I}$, is (in the basis $(\varphi_{+3/2}, \varphi_{+1/2}, \varphi_{-1/2}, \varphi_{-3/2})$):

$$\frac{8\mu_B\mu_I}{5I}|\Psi(\vec{R})|^2\Omega\left\langle\frac{1}{\rho^3}\right\rangle\begin{pmatrix} I_z & \lambda(\vec{R})\frac{I_x-iI_y}{\sqrt{3}} & 0 & 0 \\ \lambda^*(\vec{R})\frac{I_x+iI_y}{\sqrt{3}} & |\lambda(\vec{R})|^2\frac{I_z}{3} & -\frac{2}{3}|\lambda(\vec{R})|^2(I_x-iI_y) & 0 \\ 0 & -\frac{2}{3}|\lambda(\vec{R})|^2(I_x+iI_y) & -|\lambda(\vec{R})|^2\frac{I_z}{3} & \lambda^*(\vec{R})\frac{I_x-iI_y}{\sqrt{3}} \\ 0 & 0 & \lambda(\vec{R})\frac{I_x+iI_y}{\sqrt{3}} & -I_z \end{pmatrix} \quad \text{(A-11)}$$

with $\lambda(\vec{R}) = \frac{\widehat{\Psi}(\vec{R})}{\Psi(\vec{R})}$, where $\Psi(\vec{R})$ and $\widehat{\Psi}(\vec{R})$ are the envelope functions of the $\varphi_{\pm 3/2}$ and $\varphi_{\pm 1/2}$ valence electrons respectively.

**APPENDIX B**

Based on the theory of Luttinger-Kohn and Bir-Pikus, the valence band structure of a strained quantum dot can be described by the following 6 x 6 Hamiltonian :

$$\begin{pmatrix} E_{v0}-P-Q & \sqrt{2}S & -R & 0 & -S & -\sqrt{2}R \\ \sqrt{2}S^* & E_{v0}-P+Q & 0 & -R & -\sqrt{2}Q & \sqrt{3}S \\ -R^* & 0 & E_{v0}-P+Q & -\sqrt{2}S & -\sqrt{3}S^* & -\sqrt{2}Q \\ 0 & -R^* & -\sqrt{2}S^* & E_{v0}-P-Q & -\sqrt{2}R^* & S^* \\ -S^* & -\sqrt{2}Q^* & -\sqrt{3}S & -\sqrt{2}R & E_{v0}-P-\Delta_{S0} & 0 \\ -\sqrt{2}R^* & \sqrt{3}S^* & -\sqrt{2}Q^* & -S & 0 & E_{v0}-P-\Delta_{S0} \end{pmatrix} \begin{vmatrix} |\tfrac{3}{2},\tfrac{3}{2}\rangle \\ |\tfrac{3}{2},\tfrac{1}{2}\rangle \\ |\tfrac{3}{2},-\tfrac{1}{2}\rangle \\ |\tfrac{3}{2},-\tfrac{3}{2}\rangle \\ |\tfrac{1}{2},\tfrac{1}{2}\rangle \\ |\tfrac{1}{2},-\tfrac{1}{2}\rangle \end{vmatrix} \quad (B-1)$$

The valence Bloch functions are defined as:

$$u_{+3/2} = \left|\tfrac{3}{2},\tfrac{3}{2}\right\rangle = \frac{|X+iY\rangle}{\sqrt{2}}|\uparrow\rangle \quad \text{(B-2a)}$$

$$u_{+1/2} = \left|\tfrac{3}{2},\tfrac{1}{2}\right\rangle = \frac{|X+iY\rangle|\downarrow\rangle - 2|Z\rangle|\uparrow\rangle}{\sqrt{6}} \quad \text{(B-2b)}$$

$$u_{-1/2} = \left|\tfrac{3}{2},-\tfrac{1}{2}\right\rangle = \frac{|X-iY\rangle|\uparrow\rangle + 2|Z\rangle|\downarrow\rangle}{\sqrt{6}} \quad \text{(B-2c)}$$

$$u_{-3/2} = \left|\tfrac{3}{2},-\tfrac{3}{2}\right\rangle = \frac{|X-iY\rangle}{\sqrt{2}}|\downarrow\rangle \quad \text{(B-2d)}$$

$$u'_{+1/2} = \left|\tfrac{1}{2},\tfrac{1}{2}\right\rangle = \frac{|X+iY\rangle|\downarrow\rangle + |Z\rangle|\uparrow\rangle}{\sqrt{3}} \quad \text{(B-2e)}$$

$$u'_{-1/2} = \left|\tfrac{1}{2},-\tfrac{1}{2}\right\rangle = \frac{|X+iY\rangle|\uparrow\rangle - |Z\rangle|\downarrow\rangle}{\sqrt{3}} \quad \text{(B-2f)}$$

($|X\rangle$, $|Y\rangle$ and $|Z\rangle$ are orbital functions with symmetry x , y and z ; $|\uparrow\rangle$ and $|\downarrow\rangle$ are the spin components, quantified along the z-axis).

$E_{v0}$ is the $\Gamma_8$ valence band edge and is aligned relative to the valence band of the dot or matrix material, (the confinement effect is included in the spatial dependence of $E_{v0}$). $\Delta_{S0}$ is the spin-orbit split-off energy. The Hamiltonian matrix elements are given as a sum of kinetic terms and its strain counterpart:

$$P = \left(\frac{\hbar^2}{2m_0}\right)\gamma_1\left(k_x^2 + k_y^2 + k_z^2\right) - a_v\left(\varepsilon_{xx} + \varepsilon_{yy} + \varepsilon_{zz}\right) \quad \text{(B-3a)}$$

$$Q = \left(\frac{\hbar^2}{2m_0}\right)\gamma_2\left(k_x^2 + k_y^2 - 2k_z^2\right) - \frac{b_v}{2}\left(\varepsilon_{xx} + \varepsilon_{yy} - 2\varepsilon_{zz}\right) \quad \text{(B-3b)}$$

$$R = \left(\frac{\hbar^2}{2m_0}\right)\sqrt{3}\left[\gamma_2\left(k_x^2 - k_y^2\right) - 2i\gamma_3 k_x k_y\right] - \frac{\sqrt{3}}{2}b_v\left(\varepsilon_{xx} - \varepsilon_{yy}\right) + id_v\varepsilon_{xy} \quad \text{(B-3c)}$$

$$S = \left(\frac{\hbar^2}{2m_0}\right)\sqrt{6}\gamma_3 (k_x - ik_y)k_z - \frac{d_v}{\sqrt{2}}(\varepsilon_{zx} - i\varepsilon_{yz}) \quad \text{(B-3d)}$$

$\gamma_1$, $\gamma_2$ and $\gamma_3$ are the modified Luttinger parameters, $m_0$, the free electron mass, $a_v$ the valence band hydrostatic deformation potential, $b_v$ and $d_v$ the shear deformation potentials along the [001] and [111] axis.

**APPENDIX C**

In a QD, with a hole nuclear hyperfine interaction defined by the anisotropy factor $\alpha$, and in presence of a normalised magnetic field $\delta = \dfrac{B}{\Delta_0}$, the time-dependent expressions of the different spin components are respectively:

For a magnetic field applied along z,

$$R_x(t) = \int_{-\infty}^{+\infty} dx \int_{-\infty}^{+\infty} dy \int_{-\infty}^{+\infty} dz\, G(\alpha x, \alpha y, \delta + z, \tau) e^{-(x^2 + y^2 + z^2)} \quad \text{(C-1a)}$$

$$R_y(t) = \int_{-\infty}^{+\infty} dx \int_{-\infty}^{+\infty} dy \int_{-\infty}^{+\infty} dz\, F(\delta + z, \alpha x, \alpha y, \tau) e^{-(x^2 + y^2 + z^2)} \quad \text{(C-1b)}$$

$$R_z(t) = \int_{-\infty}^{+\infty} dx \int_{-\infty}^{+\infty} dy \int_{-\infty}^{+\infty} dz\, G(\delta + z, \alpha x, \alpha y, \tau) e^{-(x^2 + y^2 + z^2)} \quad \text{(C-1c)}$$

For a magnetic field applied along x,

$$R_x^1(t) = \int_{-\infty}^{+\infty} dx \int_{-\infty}^{+\infty} dy \int_{-\infty}^{+\infty} dz\, G(\delta + \alpha x, \alpha y, z, \tau) e^{-(x^2 + y^2 + z^2)} \quad \text{(C-2a)}$$

$$R_y^1(t) = \int_{-\infty}^{+\infty} dx \int_{-\infty}^{+\infty} dy \int_{-\infty}^{+\infty} dz\, F(\delta + \alpha x, \alpha y, z, \tau) e^{-(x^2 + y^2 + z^2)} = R_z^2(t) \quad \text{(C-2b)}$$

$$R_z^1(t) = \int_{-\infty}^{+\infty} dx \int_{-\infty}^{+\infty} dy \int_{-\infty}^{+\infty} dz\, G(z, \delta + \alpha x, \alpha y, \tau) e^{-(x^2 + y^2 + z^2)} \quad \text{(C-2c)}$$

$$R_y^2(t) = \int_{-\infty}^{+\infty} dx \int_{-\infty}^{+\infty} dy \int_{-\infty}^{+\infty} dz\, G(\alpha y, z, \delta + \alpha x, \tau) e^{-(x^2 + y^2 + z^2)} \quad \text{(C-2d)}$$

with $\tau = \dfrac{t}{T_{\Delta_0}}$.

The functions G and F are defined by:

$$F(a, b, c, d) = \frac{1}{\pi^{3/2}} \frac{a}{\sqrt{a^2 + b^2 + c^2}} \sin\!\left(d\sqrt{a^2 + b^2 + c^2}\right) \quad \text{(C-3a)}$$

$$G(a, b, c, d) = \frac{1}{\pi^{3/2}} \frac{a^2 + (b^2 + c^2)\cos\!\left(d\sqrt{a^2 + b^2 + c^2}\right)}{a^2 + b^2 + c^2} \quad \text{(C-3b)}$$

**FIGURE CAPTIONS**

Figure 1. Time dependence of the out-of-plane, $R_z(\tau)$, and in-plane, $R_x(\tau)$, components of the ensemble averaged spin polarization, calculated for different anisotropy factors: pure hh ($\alpha = 0$), mixed heavy/light holes ($\alpha = 0.5$), conduction electrons ($\alpha = 1$); and pure lh ($\alpha = 2$).

Figure 2. Anisotropy dependence of the steady-state values $R_x(\infty)$ and $R_z(\infty)$. For $\alpha \gg 1$ (not shown in this Figure), $R_z(\infty) \to 0$ and $R_x(\infty) \to \frac{1}{2}$.

Figure 3. Time dependence of the transverse $R_x(\tau)$, $R_y(\tau)$ and the longitudinal $R_z(\tau)$ components of the ensemble hole averaged spin. The magnetic field is applied along the z direction. Results for pure hh and lh are respectively shown on the upper (hh, $\alpha = 0$) and bottom part (lh, $\alpha = 2$) of this figure. The curves are calculated for different values of the applied magnetic field given by the reduced $\delta$ value ($\delta = B/\Delta_0$).

Figure 4. The upper part: time dependence of th transverse $R_x(\tau)$, $R_y(\tau)$ and the longitudinal $R_z(\tau)$ components of the ensemble hole averaged spin, for a magnetic field applied along the z-direction. The bottom part: time dependence of the longitudinal $R_x^1$ and transverse $R_y^1$, $R_z^1$ components of the ensemble averaged spin, for a magnetic field applied along the x direction. The curves are calculated for $\delta = 0, 0.5, 1, 2, 4$. The anisotropy factor is $\alpha=0.5$.

Figure 5. Time dependence of the longitudinal $R_x^1(\tau)$ and transverse $R_y^1(\tau)$, $R_z^1(\tau)$ components of the ensemble hole averaged spin (hh case, $\alpha = 0$). The magnetic field is applied along the x direction. The curves are calculated for different magnetic fields and represented as afunction of $\delta$ (some $\delta=4$ curves have been omitted for clarity).

Figure 6. Time dependence of the longitudinal $R_x^1(\tau)$ and transverse $R_y^1(\tau)$, $R_z^1(\tau)$ components of the ensemble hole averaged spin (lh case, $\alpha = 2$). The magnetic field is applied along the x direction. The curves are calculated for different magnetic fields and represented as a function of $\delta$ (some $\delta=4$ curves have been omitted for clarity).

Figure 7. Magnetic field dependence of (a) the steady-state value of the spin z-component $R_z(\infty)$ for an external magnetique field applied along z direction and several anisotropy factor values $\alpha = 0, 0.25, 0.5, 1, 2$; ($\delta = B_z/\Delta_0$) (b) the steady-state value of the spin z-component $R_z^1(\infty)$ for an external magnetique field applied along x direction and several anisotropy factor values $\alpha = 0, 0.25, 0.5, 1, 2$; ($\delta = B_x/\Delta_0$).

**TABLE CAPTIONS**

Table I Values of the hyperfine coupling parameter $M_j$ for an electron ($M_j=A_j$) and a heavy-hole ($M_j=C_j$) for different species j of atoms found in common III-V and II-VI compounds. a) ref [23], b) ref. [24], c) ref. [25] and d) ref. [26].

Table II. Values of the dephasing time and of the nuclear field amplitude fluctuation for a QD of typical size $N_L = 6.10^4$. The Landé factor is given for carrier; $g_h^x$ and $g_h^z$ are respectively the in-plane and the out-of-plane hh Landé factors. a) ref [27], b) ref. [21], c) ref. [28], d) ref. [19] e) ref [20] and f) ref. [29].

Table I

| species | $M_j$(eV) | | I | isotope concentration (%) |
|---|---|---|---|---|
| | Electron[a),b)] | heavy-hole [c),d)] | | |
| Ga | 38 | 3.0 | 3/2 | 100 |
| In | 56 | 4.0 | 9/2 | 100 |
| As | 46 | 4.4 | 3/2 | 100 |
| Al | | 1.2 | 5/2 | 100 |
| Cd | -30 | -3.9 | 1/2 | 25 |
| Te | -45 | -16.5 | 1/2 | 7.9 |
| Se | | 6.5 | 1/2 | 7.6 |
| S | | 1.4 | 3/2 | 0.75 |
| Zn | | 0.7 | 5/2 | 4.1 |

Table II

| QD compositon | electron $T_{\Delta_0}$ (ns) | heavy hole $T_{\Delta_0}$ (ns) | electron | | heavy hole | | | |
|---|---|---|---|---|---|---|---|---|
| | | | $\|g_e\|$ | $\Delta_0$ (mT) | $\|g_h^z\|$ | $\Delta_0$ (mT) | $\|g_h^x\|$ | $\Delta_0$ (mT) |
| InAs | 0.5 | 6.5 | 0.4[a] | 57 | 1.6[b] | 1.1 | 0.12[b] | 15 |
| GaAs | 1.2 | 12 | 0.55[c] | 17 | 2.24[c] | 0.4 | 0.09[c] | 10 |
| CdTe | 8.2 | 32 | 0.45[d] | 3.1 | 0.53[d] | 0.7 | 0.16[d] | 2.2 |
| CdSe | | 61 | 1.1[E] | | 2.5[f] | 0.07 | 0.38[f] | 2.0 |

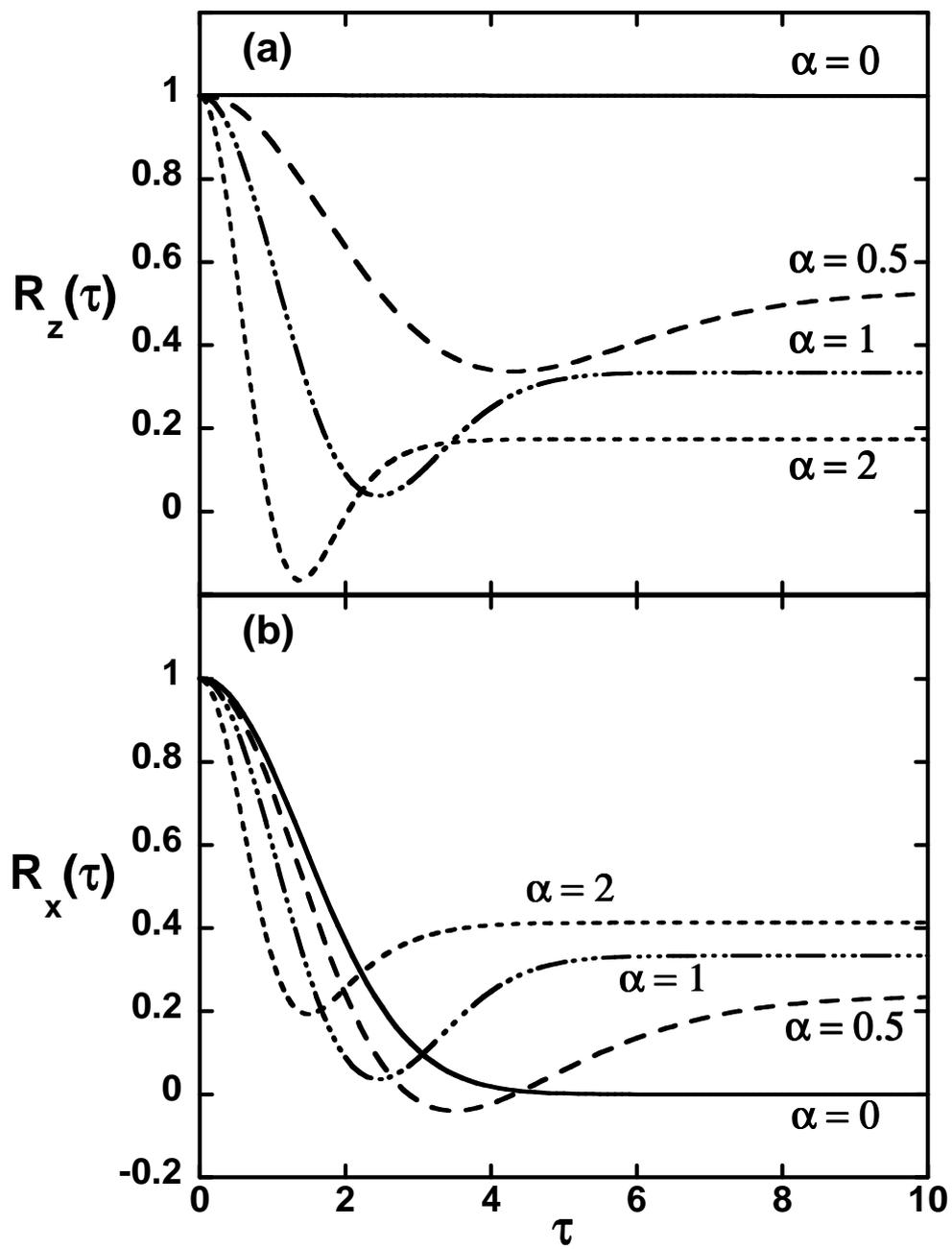

Figure 1

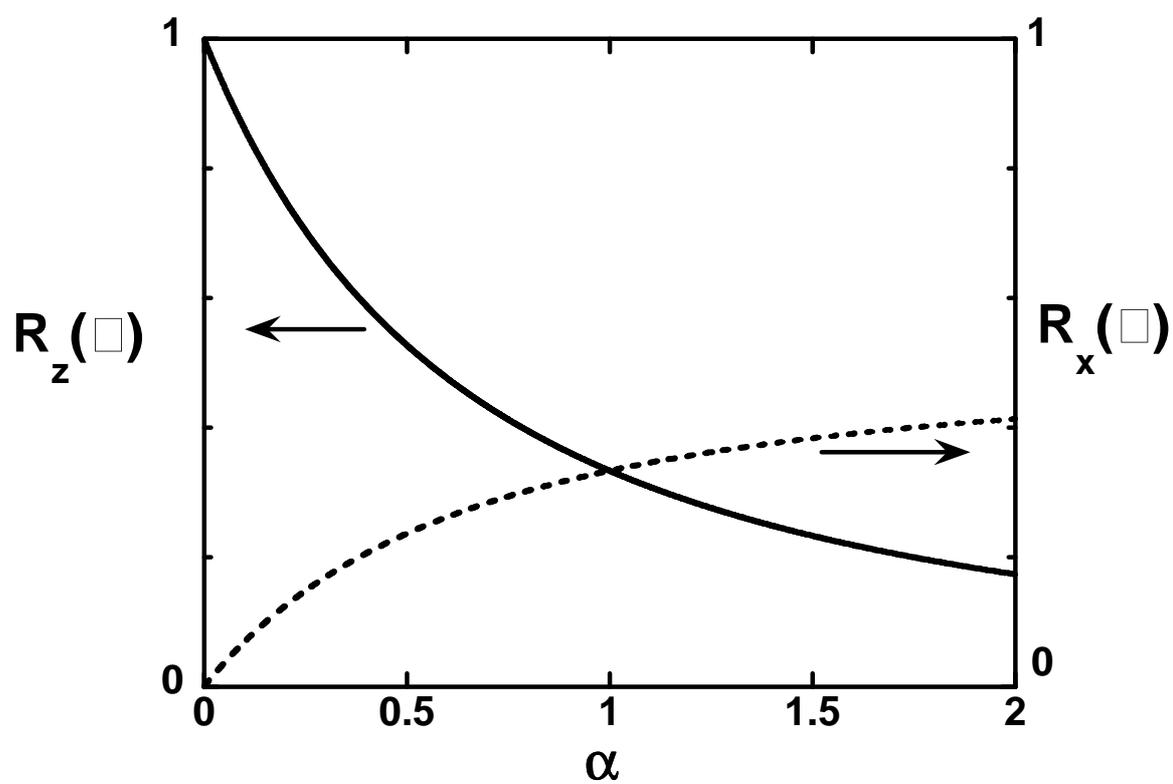

Figure 2

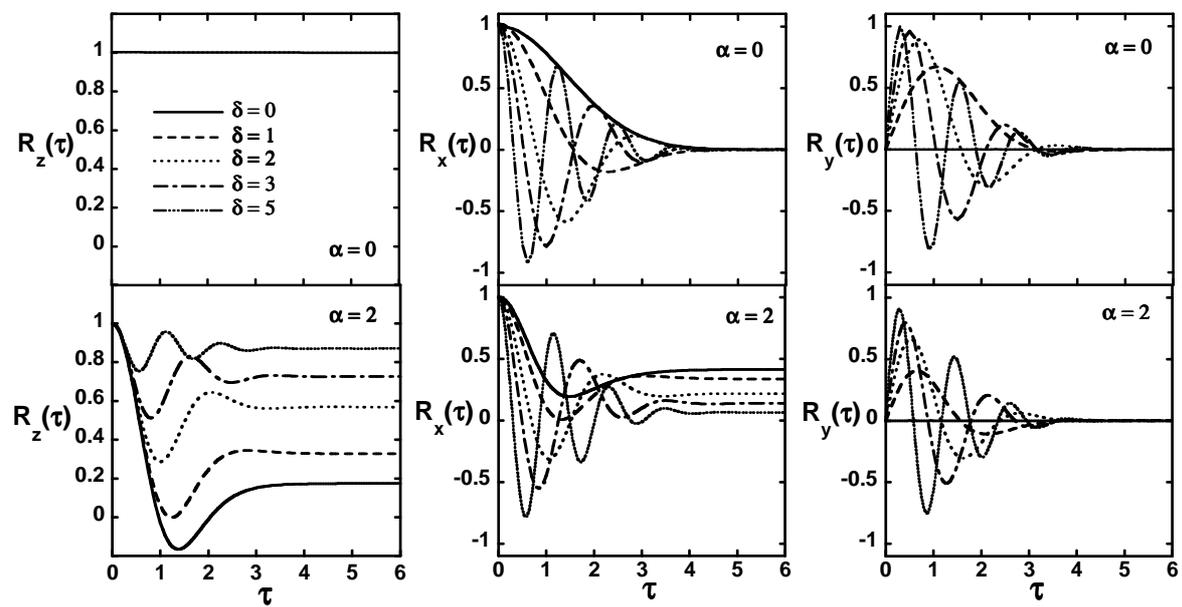

Figure 3

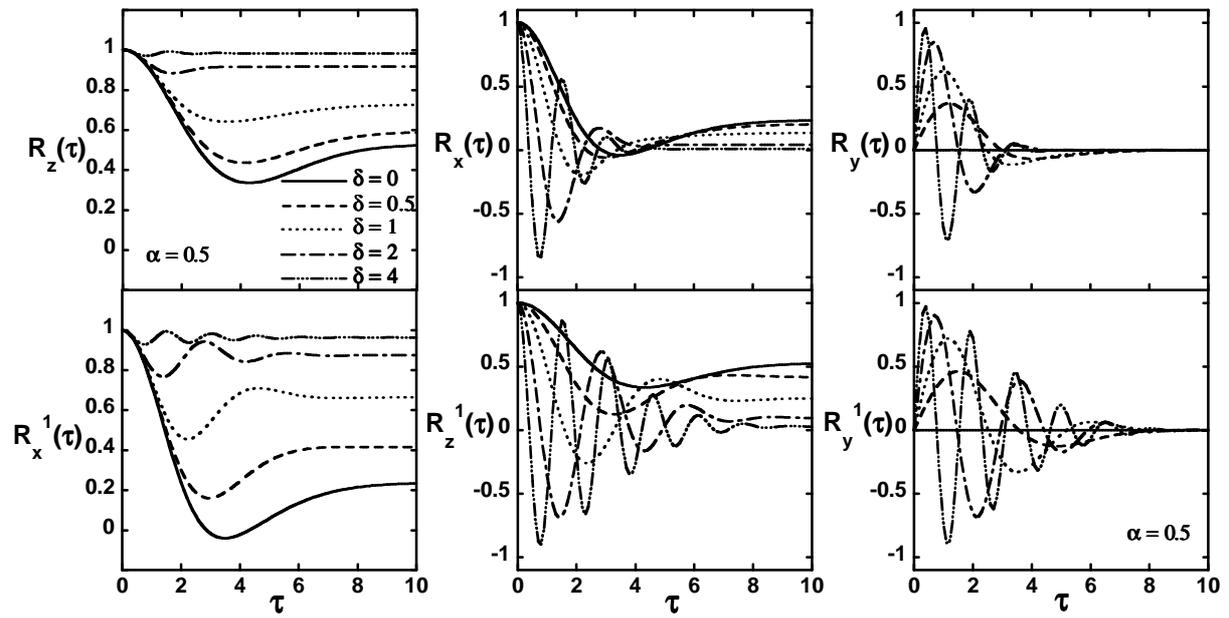

Figure 4

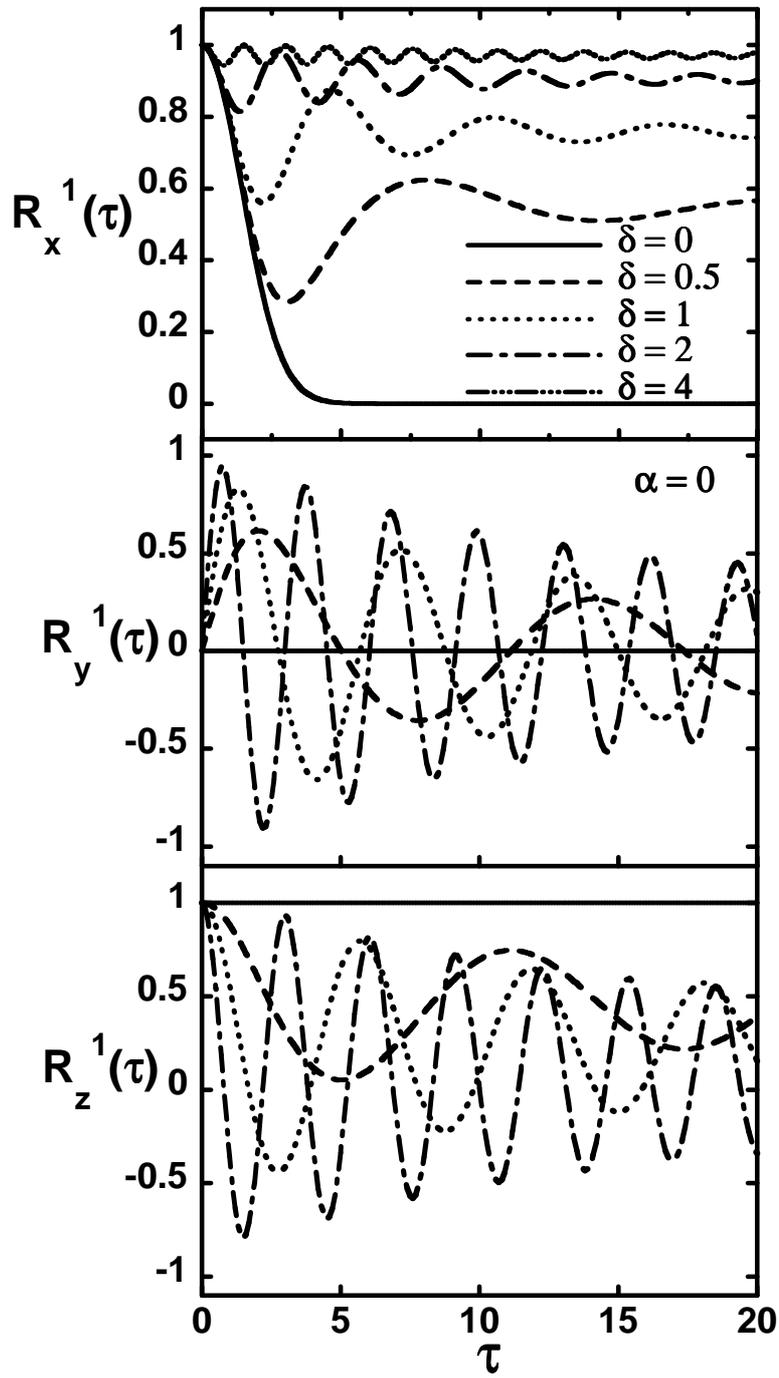

Figure 5

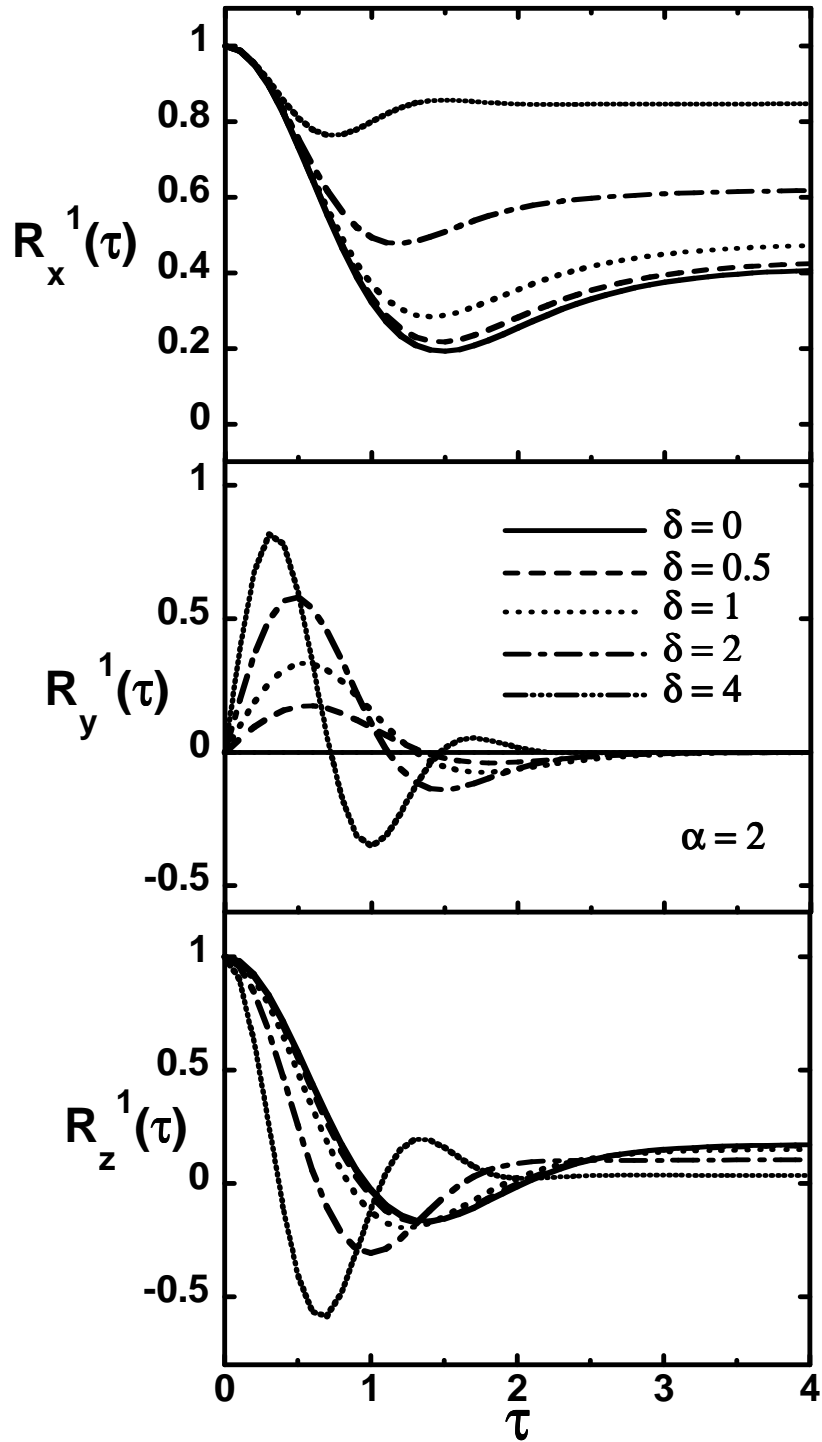

Figure 6

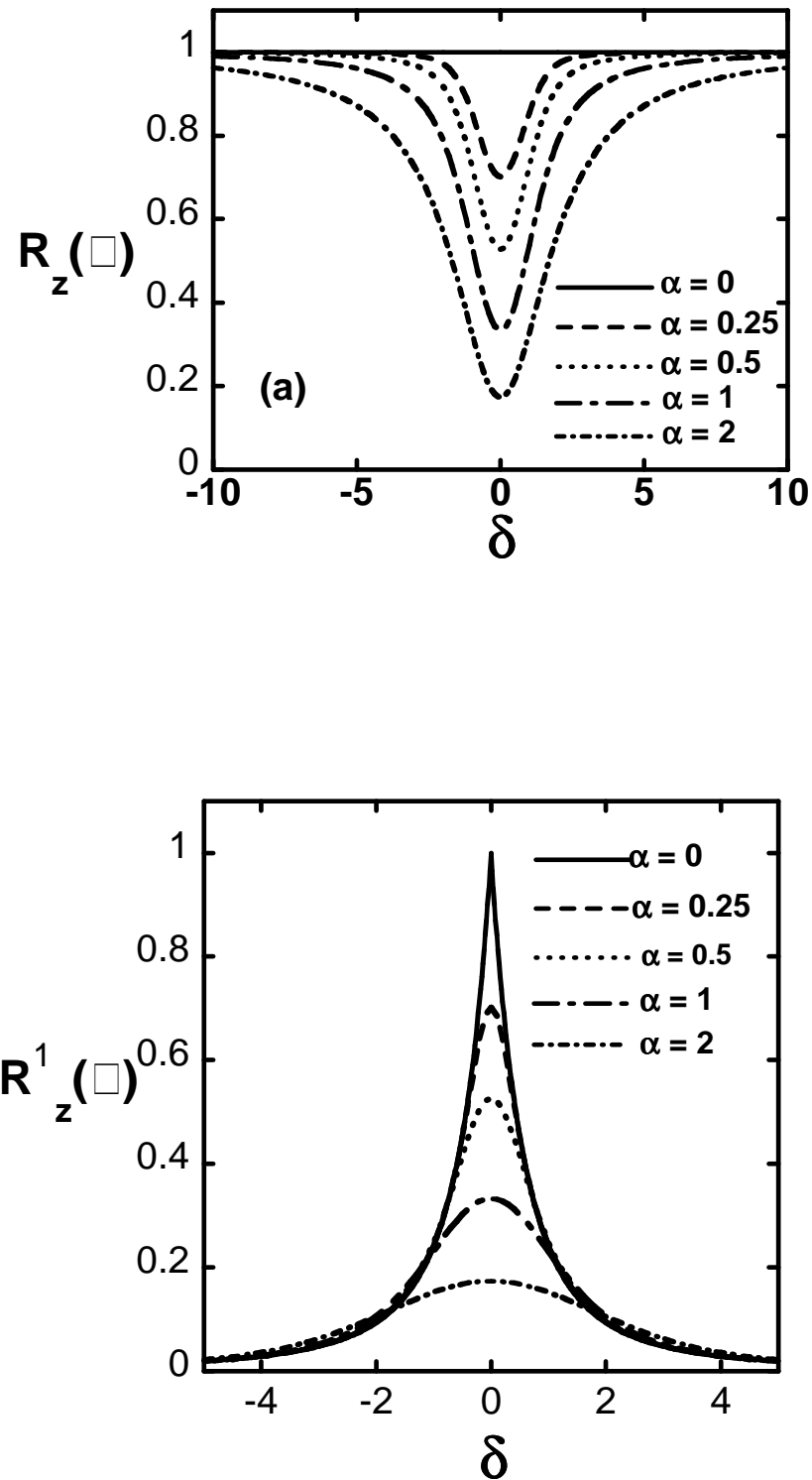

Figure 7